\documentclass{XrU2005} 
\usepackage{epsf} 
\usepackage{epsfig} 
 
\title{O Star X-ray Line Profiles Explained by Radiation Transfer in Inhomogeneous 
Stellar Wind} 
\author{L.\,M.~Oskinova} 
\author{A.~Feldmeier}
\author{W.-R.~Hamann}
\affil{Universit{\"a}t Potsdam, Astrophysik, 14469 Potsdam, Germany}

\begin{document} 
 
 
\maketitle 
 
\begin{abstract} 
It is commonly adopted that X-rays from O stars are produced deep inside
the stellar wind, and transported outwards through the bulk of the 
expanding matter which attenuates the radiation and affects the shape of
emission line profiles. The ability of Chandra and XMM-Newton to resolve
these lines spectroscopically provided a stringent test for the theory of
X-ray production. It turned out that none of the existing models was
able to reproduce the observations consistently. The major caveat of these
models was the underlying assumption of a smooth stellar wind. Motivated
by the various observational evidence that the stellar winds are in fact
structured, we present a 2-D model of a stochastic, inhomogeneous wind.
The X-ray radiative transfer is derived for such media.  It is shown that
profiles from a clumped wind differ drastically from those predicted by 
conventional homogeneous models. We review the up-to-date observations 
of X-ray line profiles from stellar winds and present line fits obtained 
from the inhomogeneous wind model. The necessity to account for inhomogeneities 
in calculating the X-ray transport in massive star winds, including for HMXB 
is highlighted.
\end{abstract} 
 
\section{Clumped winds from O-type stars} 

Young and massive O-type stars possess strong stellar winds. The
winds are fast, with typical velocities up to 2500 km/s, and dense, with 
mass-loss rates  $\dot{M} \sim 10^{-7} M_{\odot}/{\rm yr}$. The driving mechanism 
for the mass-loss from OB stars has been identified with radiation pressure on 
spectral lines. A corresponding theory was developed by \citet{CAK75}. After consequent 
improvements, this theory correctly predicts the observed mass-loss rates and wind 
velocities  \citep{Pldrch86}. However, it was pointed out early \citep{Lucy70}, and 
later further investigated \citep{OR84}, that the stationary solution for a 
line-driven wind is unstable; small perturbation grow quickly and result in strong 
shocks. The most detailed hydrodynamic modeling of the line driven instability was 
presented by \citet{AF97}. These calculations show that the winds are strongly 
inhomogeneous with large density, velocity and temperature variations. The density 
inhomogeneity is commonly referred to as wind clumping. \citet{RO05} studied the 
1-D evolution of instability-generated structures in the winds. They have demonstrated 
that the winds are essentially clumped out to distances of 1000\,$R_*$.  

The theoretical prediction of the wind clumping is confirmed observationally. 
\citet{Bouret05} conducted a quantitative analysis of the far-ultraviolet spectrum 
of two Galactic O stars using the last generation NLTE stellar atmosphere  codes. 
Their study provided a strong evidence of wind clumping in O stars. It was pointed 
out that accounting for wind clumping is essential when determining the wind 
properties of O stars.

One of the most significant consequences of the clumping is the reduction of
empirically derived mass-loss rates  by at least a factor of 3 \citep{hk98,Bouret05}. 
The stellar mass-loss rate determines the wind density, and therefore the wind 
opacity. Studying the effects of absorption in the observed X-ray spectra allows 
to probe the wind opacity. Analysis of the X-ray emission from individual stellar 
winds in the O stars $\delta$~Ori \citep{mil02} and $\zeta$~Pup \citep{kr03} have 
shown that the attenuation by the stellar wind is significantly smaller than expected 
from standard homogeneous wind models. 

%%%%%%%%%%%%%%%%%%%%%%%%%%%%%%%%  FIGURE 1 %%%%%%%%%%%%%%%%%%%%%%%%%%%%%%%%%%%%
\begin{figure} 
\centering 
\epsfig{file=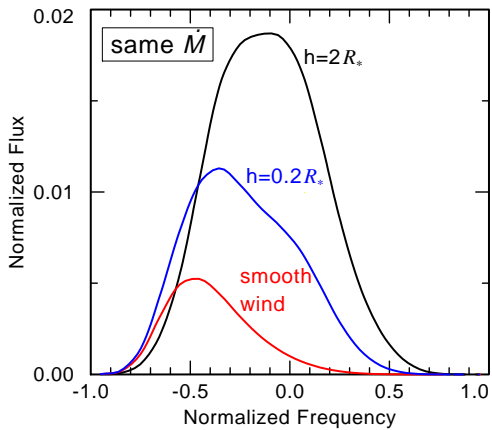,width=0.8\linewidth}  
\caption{Comparison between example line profiles emerging from a clumped and a 
smooth wind of the same mass-loss rate, velocity law ($\beta=1.5$), and equal X-ray 
emissivity. Frequency is measured relative to the line center and in Doppler units 
referring to the terminal wind velocity $v_\infty$. The average separation between 
clumps $h$ is indicated.
\label{fig:lin}} 
\end{figure} 
%%%%%%%%%%%%%%%%%%%%%%%%%%%%%%%%%%%%%%%%%%%%%%%%%%%%%%%%%%%%%%%%%%%%%%%%%%%%%%%

Similar conclusions are reached from the analysis of X-rays from colliding wind 
binaries (CWB). Such systems consist of two massive early type stars, where each 
component has a strong wind. The copious X-ray emission is produced in the zone where 
the stellar winds collide. At certain orbital phases the X-rays formed in the 
colliding wind zone travel towards an observer through the bulk of the stellar wind 
of one companion. Deriving the absorbing column density from X-ray spectroscopy 
constrains the wind density. Mass-loss rates can be inferred and compared with the 
models. An analysis of XMM-Newton observations of the massive binary $\gamma^2$\,Vel 
was presented by \citet{sch04}. They deduced the  column density of absorbing material 
from fits of the X-ray spectra and showed that the observed attenuation of X-rays 
is much weaker than expected from smooth stellar wind models. To reconcile theory with 
observations, \citet{sch04} suggest that the ratio of clump density to smooth wind 
density has to be high, and the volume filling factor of the clumps has to be small. 
Significantly, similar conclusions were reached from  Chandra and 
RXTE observations of WR\,140 \citep{pol05}. Alike $\gamma^2$\,Vel, the column density 
expected from the stellar atmosphere models that account for the clumping in first 
approximation only, is a factor of four higher than the column density inferred from 
the X-ray spectrum analysis. 

Spectacular evidence of wind clumping comes from X-ray spectroscopy of high-mass 
X-ray binaries (HMXB). In some of these systems a neutron star (NS) is on a close 
orbit deeply inside the stellar wind of the massive star. The X-ray emission with 
a power law spectrum is produced as a result of Bondi-Hoyle accretion of the stellar 
wind onto the NS. These X-rays photoionize the stellar wind. The resulting X-ray
spectrum shows a large variety of emission features, including lines from H-like 
and He-like ions and a number of fluorescent emission lines. \citet{sako03} reviewed 
spectroscopic results obtained by X-ray observatories for several wind-fed HMXBs. 
They conclude that the observed spectra can be explained only as originating in a 
clumped stellar wind where cool dense blobs are embedded in rarefied  photoionized 
gas. \citet{vdm05} studied stochastic variability of the X-ray light curve in 
4U~1700-37 and its X-ray spectra. They shown that the feeding of the NS by a strongly 
clumped stellar wind is consistent with the observed temporal and spectral variability.

X-ray emission is also intrinsic for clumped stellar winds of single stars. \citet{AF97} 
performed hydrodynamic simulations of an O star wind. They showed that the de-shadowing 
instability leads to strong gas compression resulting in dense cool shell fragments. 
The space between fragments is essentially void, but at the outer side of the dense 
shells exist extended gas reservoirs.  Small gas cloudlets are ablated from these 
reservoirs, and accelerated to high speed by radiation pressure. Propagating through 
an almost perfect vacuum, they catch up with the next outer shell and ram into it. In 
this collision, the gas parcels are heated and emit thermal X-rays. The X-ray emission 
ceases when the wind reaches its terminal speed. In contrast, the cool fragments are 
maintained out to large distances, despite their expansion due to the internal pressure.  
Thus there are two disjunctive structural wind components -- hot gas parcels that emit 
X-rays and highly compressed cool fragments that attenuate this radiation. 

\section{Transfer of X-rays in a clumped stellar wind}

%%%%%%%%%%%%%%%%%%%%%%%%%%%%%%%%  FIGURE 2 %%%%%%%%%%%%%%%%%%%%%%%%%%%%%%%%%%%%
\begin{figure*} 
\centering 
\epsfig{file=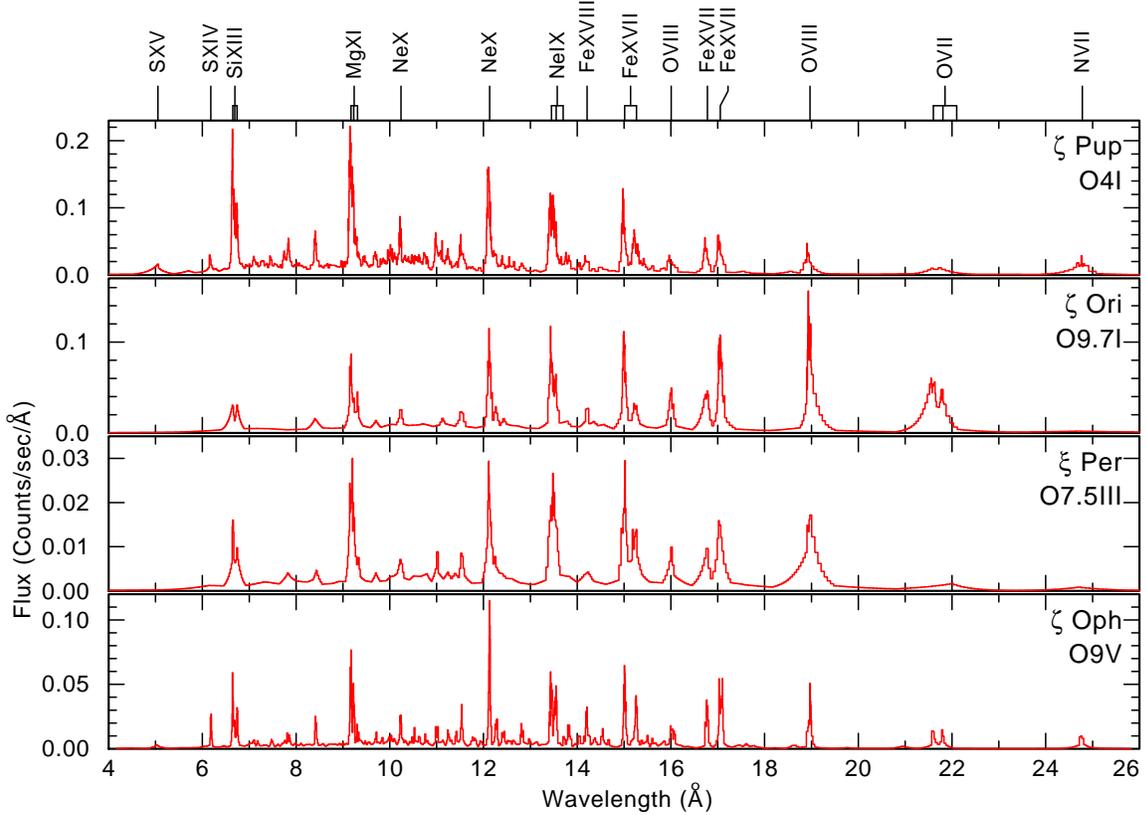, width=10.8cm, angle=-90}  
\caption{De-reddened HETGS spectra of prominent O stars. All stars except $\zeta$~Ori 
are evolved runaway stars, as reflected by the enhanced N/O ratio. 
\label{fig:der}} 
\end{figure*} 
%%%%%%%%%%%%%%%%%%%%%%%%%%%%%%%%%%%%%%%%%%%%%%%%%%%%%%%%%%%%%%%%%%%%%%%%%%%%%%%

Wind inhomogeneity alters the radiative transfer significantly. We have studied 
the effects of wind fragmentation on  X-ray line formation analytically in the limit 
of infinitely many stochastically distributed fragments  \citep{feld03}, and 
numerically for more realistic stellar winds \citep{osk04}. The latter work describes 
the model and the code used in the present paper to reproduce the observed X-ray 
emission line profiles. It is assumed that X-rays are emitted by a large number of 
stochastically distributed hot gas parcels within the wind acceleration zone. These 
gas parcels are permeated with the numerous aligned cool shell fragments (i.e. clumps). 
The fragments are also distributed stochastically and propagate, maintaining their 
solid angle, outwards with $v(r)$. They persist till large distances from the stellar 
surface. 

Because the average stellar mass-loss rate is constant, the mass confined within each 
clump is, on average, the total mass of the stellar wind divided by the  number of 
shell fragments. The mass-loss rate $\dot{M}$ is fixed for a specific star from 
UV/optical spectral analysis. We employ the average number density of the fragments, 
$n(r)$,  as a free parameter. The fragments are radially compressed and therefore 
aligned. Thus the optical depth, $\tau_{\rm sh}(E,\,{\vartheta_{\rm sh}})$, of each 
individual shell fragment depends on its mass, the mass absorption coefficient 
at energy $E$, and the intersection angle between the shell and the line of sight 
($\vartheta_{\rm sh}$).

The fate of each X-ray photon emitted deep inside the wind and propagating outwards 
depends on whether it escapes from the wind without encountering any shell fragment, 
or it accidentally hits a fragment. The chance to hit a shell is smaller when the 
average separation between shells, $h$, is larger. The average separation $h$ depends
on the number density of the fragments and their cross-section: 
$h=\left<\,n(r)\sigma\,\right> ^{-1}$.   

The probability that the photon that has hit a shell is absorbed  is 
$1-\exp(-\tau_{\rm sh})$. Therefore the {\em effective opacity $\bar{\tau}$ } in the 
stellar wind depends on the number density of shells $n(r)$, the cross section of the 
shells $\sigma$, and the probability of a photon to be absorbed in a shell. The optical 
depth along a photon path through the clumpy wind is
\begin{equation}
\bar{\tau}=\int n(r)\,\sigma\,(1-e^{-\tau_{\rm sh}(E)}) dr
\label{eq:opa}
\end{equation}  
Importantly, in the limit of a very large number of fragments, Eq.\,(\ref{eq:opa})
recovers the formula for the optical depth  in a smooth wind \citep{osk04}.   

Assume now that radiation of intensity  $I_0$ is emitted somewhere in the wind. 
After passing through the wind this intensity will be reduced to
\begin{equation} 
I=I_0 e^{-\bar{\tau}}
\label{eq:line}
\end{equation}
From Eqs.\,({\ref{eq:opa},\,\ref{eq:line}}) one can see that for optically thick 
clumps ($\tau_{\rm sh}\gg 1$) the effective optical depth $\bar{\tau}$ is independent 
of the energy. The opacity in the clumped wind is effectively grey. 

Our modeling shows that  the wind fragmentation drastically reduces the opacity of 
the wind. Therefore X-rays produced deep inside the wind can effectively escape,
which would be totally absorbed in a homogeneous wind of the same mass-loss rate. 
A comparison between predicted line profiles emerging from clumped and smooth winds of 
the same mass-loss rate is shown in Fig.\ref{fig:lin}. 

%%%%%%%%%%%%%%%%%%%%%%%%%%%%%%%%  FIGURE 3 %%%%%%%%%%%%%%%%%%%%%%%%%%%%%%%%%%%%
\begin{figure*} 
\centering 
\epsfig{file=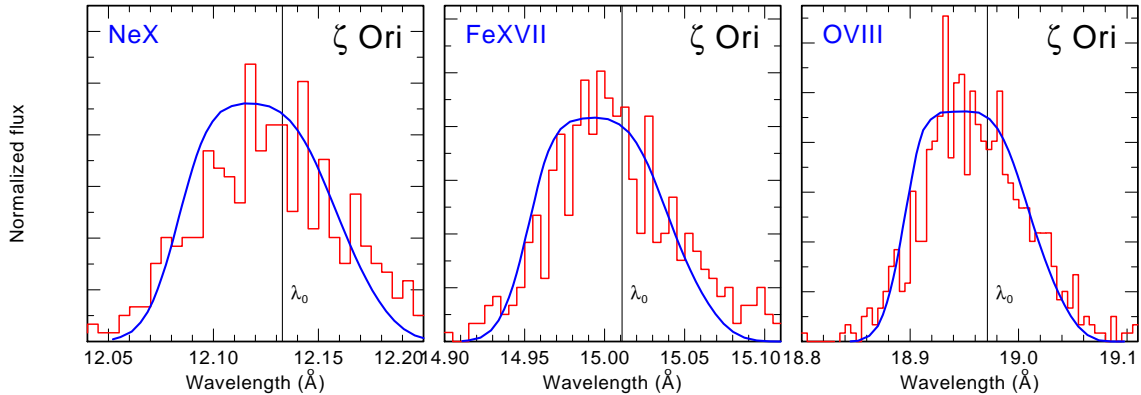, height=15cm, angle=-90}  
\caption{Thin lines: co-added meg$\pm 1$ lines observed in the spectrum of 
$\zeta$~Ori. Thick lines: model line profiles. The rest wavelength corrected 
for the stellar radial velocity is indicated by $\lambda_0$. Model line profiles 
are calculated using $h\approx 1\,R_*$,~$\beta=3$ and the appropriate for this star 
$\dot{M}$ and $v_\infty$.  
\label{fig:zori}} 
\end{figure*} 
%%%%%%%%%%%%%%%%%%%%%%%%%%%%%%%%%%%%%%%%%%%%%%%%%%%%%%%%%%%%%%%%%%%%%%%%%%%%%%%

The line width is determined by the velocity field of the hot plasma. It is commonly 
accepted to approximate the velocity field in the stellar wind by a ``$\beta$-law'' 
of the form 
\begin{equation}
v(r)=v_\infty \left(1-\frac{1}{r}\right)^\beta, 
\label{eq:betalaw}
\end{equation}
where the distance $r$ is normalized to the stellar radius. The terminal velocity 
$v_\infty$ is deduced by analyzing UV/optical spectra. Hence, the free parameter 
to describe the width of an X-ray line is $\beta$.

Intrinsic line intensity $I_0$ can be calculated from hydrodynamic wind models but it 
is not done yet. To compare the model and the observed line profiles we normalize the 
flux in the model line to the observed one. Thus only two free parameters are required  
to fit the observed line: the average separation between clumps $h$, which defines the 
attenuation; and $\beta$, which defines the line broadening.

\section{Grating observations of O stars } 

We have selected Chandra HETGS observations of prominent O stars to fit the observed 
line profiles in the framework of a clumped wind model. The de-reddened spectra of 
selected stars are shown in Figure~\ref{fig:der}. We have chosen three stars that are 
most likely single, and therefore their X-ray emission can be solely attributed to 
the stellar wind itself.  The probability of a massive star being single is significantly 
higher among runaway stars and three stars in our sample are runaway stars that have not 
shown indications of binarity. The forth star, $\zeta$~Ori, is a known binary. We 
include $\zeta$~Ori in the sample because the initial interpretation of the observed 
emission lines from this star caused doubts in the validity of the shock model of 
X-ray production \citep{wal01}. The wind parameters of the stars are listed in Table~1. 

%%%%%%%%%%%%%%%%%%%%%%%%%%%%%%%%%%% TABLE 1 %%%%%%%%%%%%%%%%%%%%%%%%%%%%%%
\begin{table} 
  \begin{center}  
   \caption{Stellar wind parameters from \citet{Rep04} and \citet{lam99}}\vspace{1em} 
    \renewcommand{\arraystretch}{1.2} 
    \begin{tabular}[h]{lcc} 
      \hline 
      Name  & $v_\infty$ [km/s] & $\dot{M}$ [$M_\odot {\rm yr}^{-1}$]\\ 
      \hline 
      $\zeta$~Pup  & 2250 & $ 9.0\times 10^{-6}$    \\ 
      $\zeta$~Ori  & 2100 & $ 2.5\times 10^{-6}$           \\ 
      $\xi$~Per    & 2450 & $ 1.8\times 10^{-6}$\\ 
      $\zeta$~Oph  & 1550 & $\leq 1.8\times 10^{-7}$ \\
      \hline \\ 
      \end{tabular} 
    \label{tab:table} 
  \end{center} 
\end{table} 
%%%%%%%%%%%%%%%%%%%%%%%%%%%%%%%%%%%%%%%%%%%%%%%%%%%%%%%%%%%%%%%%%%%%%%%%%%% 

The observed spectra in the $4-24$\,\AA~ band were fitted using the {\em xspec} package. 
The spectrum of  $\zeta$~Pup has the highest  S/N. A satisfactory spectral fit for 
this star can be achieved only by using a model that accounts for the continuum emission
as well as for the lines. The {\em apec} model is capable to reproduce the X-ray  emission 
lines in the $\zeta$~Pup spectrum. The continuum can be fitted by thermal bremsstrahlung. 
Importantly, the temperatures obtained from the emission lines and from the continuum 
are identical, $kT_{\rm X}\approx 0.6$\,keV, and in good agreement with the predictions 
of the wind shock model. 

The plasma in hot star winds is not expected to be isothermal. However, the strong shock 
condition implies that the maximum temperature of the shock material scales with the 
square of the preshock  wind velocity. Constraining the temperature of hot gas in 
$\zeta$~Oph provides an interesting test, because this star has the slowest stellar
wind. It is expected that its X-ray plasma temperature should be a factor of 
two lower than in $\zeta$~Pup. To estimate temperatures we used H-like to He-like ion 
line ratios of Mg and Si. Our estimates show that the temperature in $\zeta$~Oph is 
indeed lower than in $\zeta$~Pup. A temperature less than 0.4\,keV is also consistent 
with the results of fitting the $\zeta$~Oph spectrum with {\em xspec}.

The strongest emission lines observed in the spectra of our sample stars are shown in 
Figs.\,3,\,4,\,5 and 6. Lines in all spectra are broad, blueshifted and symmetric. 
The Ne\,{\sc x} line is a  doublet ($\lambda_{0_1,0_2}=12.132,\,12.138$\,\AA). 
\citet{kr03} provided a detailed analysis of emission lines in $\zeta$~Pup using the 
smooth wind model. They concluded that the lines are broadened to the terminal 
wind velocity and have similar shapes across the spectrum. 

\section{Model fitting of observed lines}

A detailed modeling  of the X-ray spectra and emission line profiles can 
be done only by means of full stellar atmosphere models.  Conventional ``standard'' 
models, e.g.  {\em apec}, strictly speaking are not adequate for the fast moving 
stellar winds, and cannot predict expected line profiles. The approach of \citet{kr03} 
was to normalize the model line flux to the observed one. The line profile fitting was 
used to constrain the wind optical depth at the line wavelength. This is possible because 
in the smooth wind approximation, the shape of emission line is sensitive to the amount 
of material on the line of sight. The more skewed the line is, the more opaque is the wind 
\citep{ig01}.    

%%%%%%%%%%%%%%%%%%%%%%%%%%%%%%%%  FIGURE 5 %%%%%%%%%%%%%%%%%%%%%%%%%%%%%%%%%%%%
\begin{figure} 
\centering 
\epsfig{file=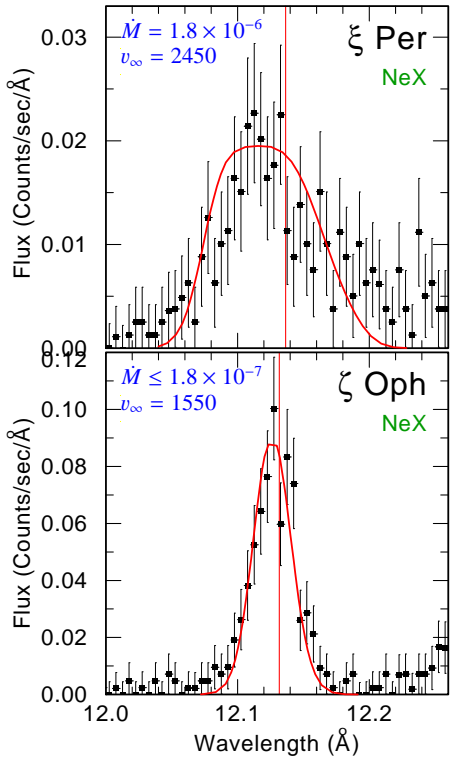,height=10cm}
\caption{Observed and model Ne{\sc x} line in $\xi$~Per and $\zeta$~Oph. 
Model line profiles are calculated using $h\approx 1\,R_*$ and $\beta=3$ for both 
stars. $\dot{M}$ and $v_\infty$ are from Table~1.  
\label{fig:com1}} 
\end{figure} 
%%%%%%%%%%%%%%%%%%%%%%%%%%%%%%%%%%%%%%%%%%%%%%%%%%%%%%%%%%%%%%%%%%%%%%%%%%%%%%%
  
This does not hold in a clumped wind. The shape of an emission line is sensitive 
to the spatial distribution of the clumps in the winds. When a large number of clumps 
is effectively black, the radiation can still escape between them. Thus the  average  
separation between clumps is the key parameter determining how much X-ray radiation 
escapes. When clumps are compressed in radial direction, the line profile 
is distinctly symmetric although its maximum is blueshifted as can be seen in Fig.\,1. 
When the separation between clumps is small, the wind is nearly homogeneous and emergent 
emission lines have skewed shapes characteristic for a smooth wind. However, if the 
separation between clumps is large enough, the lines shape becomes symmetric. Therefore, 
the average clump separation can be constrained by fitting the observed line profiles.
%
%%%%%%%%%%%%%%%%%%%%%%%%%%%%%%%%  FIGURE 6 %%%%%%%%%%%%%%%%%%%%%%%%%%%%%%%%%%%%
\begin{figure} 
\centering 
\epsfig{file=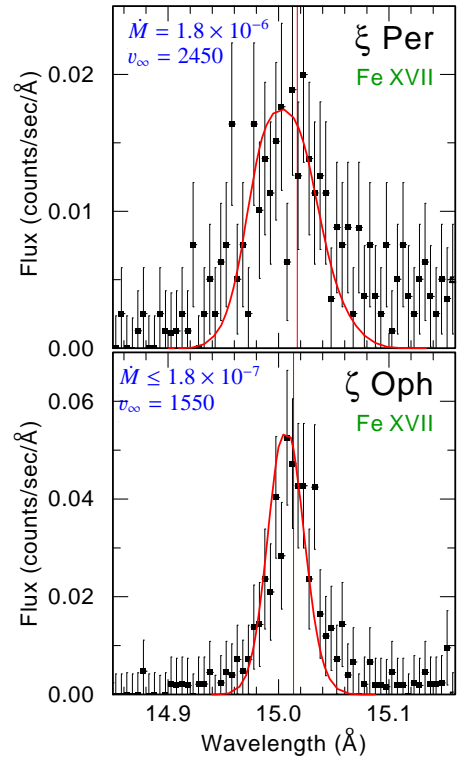,height=10cm}
\caption{Observed and modeled Fe{\sc xvii} line in $\xi$~Per and $\zeta$~Oph. 
Model parameters the same as in Fig.\,4.
\label{fig:com2}} 
\end{figure} 
%%%%%%%%%%%%%%%%%%%%%%%%%%%%%%%%%%%%%%%%%%%%%%%%%%%%%%%%%%%%%%%%%%%%%%%%%%%%%%%
%
Shown in Fig.\,\ref{fig:zori} are the lines observed in the spectrum of $\zeta$~Ori.
As can be seen, the observed lines are broad, symmetric and blueshifted. The lines 
are very well reproduced by our fragmented wind model.  Line profiles
resolved by Chandra in the spectra of $\xi$~Per and $\zeta$~Oph and model fits 
for these lines are shown in Figs.\,\ref{fig:com1} and \ref{fig:com2}. All model 
fits are produced assuming that the average separation between the shell fragments 
is of the order of the stellar radius. 

The distance from the stellar core where X-ray emission originates can be estimated 
from line ratios of He-like ions. The strong UV field of an O star causes radiative 
de-population of metastable $^3$S level, weakening forbidden (F) line in favor 
of the recombination (I) line \citep{por01}. Therefore ratio F/I is a diagnostic of 
UV field. The UV field dilutes with radius, thus the ratio F/I provides information 
about the distance from the stellar photosphere. The values of F/I obtained for 
different ions are similar. For instance, our analysis of Mg\,{\sc xi} and Si\,{\sc xiii}
in $\zeta$~Oph indicates that the plasma is located between 1.8 and 6\,$R_*$ from the 
stellar core. 

When the location of the hot plasma is constrained and the terminal velocity of the 
wind is known, parameter $\beta$ can be varied to adjust the model line width. As can 
be seen in Figs.\,\ref{fig:com1} and \ref{fig:com2}, the width of lines differs 
significantly between stars. Lines observed in $\xi$~Per, the star with the fastest
wind, are much broader than the lines in $\zeta$~Oph, where the stellar wind is slowest.  

\begin{figure} 
\centering 
\epsfig{file=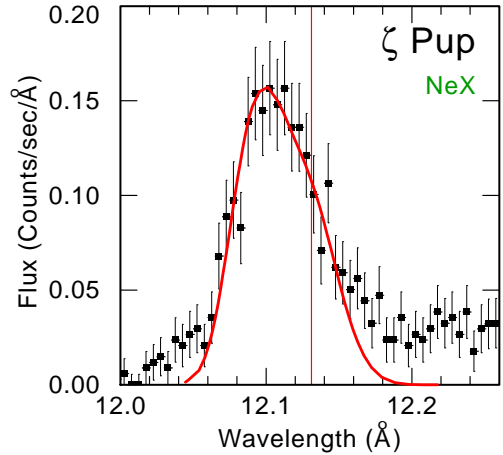,width=0.8\linewidth}  
\caption{Observed and model Ne\,{\sc x} line in $\zeta$~Pup.
Model line profile is calculated using $h\approx 0.2\,R_*$,~$\beta=1.5$ and the 
appropiate for this star $\dot{M}$ and $v_\infty$.  
\label{fig:pup}} 
\end{figure}

The model lines shown in Figs.\,\ref{fig:zori},\,\ref{fig:com1} and \ref{fig:com2} 
required rather large values of $\beta\geq 2$, indicating slow wind acceleration.  
This $\beta$ is much higher than the usual $\beta\approx 0.8$ inferred 
from analyses of UV/optical spectra and predicted from theoretical wind models 
\citep{Rep04}. The lines in the $\zeta$~Pup spectrum, fitted with smaller values of
$\beta=1.5$, are shown in Fig.\,\ref{fig:pup}. 

\section{Wind clumping in HMXBs}

%%%%%%%%%%%%%%%%%%%%%%%%%%%%%%%%  FIGURE 6 %%%%%%%%%%%%%%%%%%%%%%%%%%%%%%%%%%%%
%%%%%%%%%%%%%%%%%%%%%%%%%%%%%%%%%%%%%%%%%%%%%%%%%%%%%%%%%%%%%%%%%%%%%%%%%%%%%%%

The effects of wind clumping are important for the interpretation of the X-ray 
spectra from the wind-fed  HMXBs. Spectra obtained at orbital phases near X-ray 
eclipse are dominated by emission reprocessed in the stellar wind. The X-rays 
from the accretion onto the compact object are scattered in the wind. The Compton 
optical depth scales with density, and the effects of clumping should be included 
in the modeling of the scattering in the wind. In addition, an emission line spectrum
is emerging from the photoionizaed stellar wind near the compact object. On their way 
to the observer, X-rays are attenuated in the stellar wind. Clumping 
reduces the absorption and makes it less wavelength dependent.

Clumping also affects the size of the photoionized region. Its radius is determined 
by the balance of ionization and recombination. The ionization rate scales with density, 
but the recombination rate scales with density squared. Therefore, if the wind consists 
of numerous but optically thin clumps, the radius of the photoinized region would be 
reduced compared to a homogeneous wind. If, however, the wind is compressed in a 
relatively small number of optically thick clumps, then the photionized region becomes 
more extended compared to the the smooth wind case (but with recombined clumps embedded). 
Hence, the radius of the photoionized region is determined by the clump filling factor 
and the density of the clumps.           

\section{Conclusions}

\noindent
There are firm observational evidences of stellar winds being strongly 
inhomogeneous.

\noindent
Clumping strongly reduces the opacity of the stellar wind for the X-ray 
emisison.

\noindent
Absorption in a clumped wind is effectively independent of wavelength.

\noindent
The line profiles observed in  X-ray spectra of O stars are broad, 
symmetric and blueshifted.

\noindent
O star X-ray line profiles are explained by radiation transfer in an 
inhomogeneous stellar wind.

\end{document}